\documentclass[aps,prd,showpacs,preprintnumbers]{revtex4}
\usepackage{graphicx}
\usepackage{epsfig}
\usepackage{amsmath}
\usepackage{subfigure}

\def \beq{\begin{equation}}
\def \eeq{\end{equation}}
\def \beqa{\begin{eqnarray}}
\def \eeqa{\end{eqnarray}}
\def \beal{\begin{align}}
\def \enal{\end{align}}
\begin{document}

\title{Low-lying Scalars in an extended Linear $\sigma$ Model}

\bigskip
\bigskip
\author{Tamal K. Mukherjee$^{1,2}$}
\email{mukherjee@ihep.ac.cn}
\author{Mei Huang$^{1,2}$}
\email{huangm@ihep.ac.cn}
\author{Qi-Shu Yan$^3$}
\email{yanqishu@gucas.ac.cn}
\affiliation{$^1$  {Institute of High Energy Physics, Chinese Academy of
Sciences, Beijing, China} \\
$^2$ {Theoretical Physics Center for Science Facilities, Chinese Academy of Sciences, \\
Yuquan Road 19B, 100049, Beijing, China} \\
$^3$ College of Physics Sciences, Graduate University of Chinese Academy of Sciences,
Beijing 100039, P.R China\\}
\date{\today }
\bigskip

\begin{abstract}
We formulate an extended linear $\sigma$ model of a quarkonia nonet and a
tetraquark nonet as well as a complex iso-singlet (glueball) by
virtue of chiral symmetry $SU_L(3) \times SU_R(3)$ and $U_A(1)$
symmetry. In the linear realization formalism, 
we study the mass spectra and components of the low-lying scalars and pseudo scalars in this model. The mass matrices for physical staes are obtained and
the glueball candidates are examined. We find that
the model can accommodate the mass spectra of low-lying states quite well. 
Our fits indicate that the most glueball like scalar should be 
$2$ GeV or higher while the glueball pseudoscalar is $\eta(1756)$.
We also examine the parameter region where the lightest iso-scalar $f_0(600)$ can
be the glueball and quarkonia dominant but find such a parameter region may be confronted with 
the problem of the unbounded vacuum from below.
\end{abstract}

\maketitle

\section{Introduction}

The pseudoscalar, vector and axial-vector as well as tensor mesons of
light quarks have been well-understood in the naive quark model in terms of the chiral symmetry. 
Despite of its success, the naive quark model can not explain 
the scalar meson sector, which have the same quantum numbers
as the vacuum. There are about 19 states which are twice more than the expected ${\bar q}q$
nonet as in vector and tensor sectors, while the mass and decay pattern of these low-lying scalars are different 
from the expectation of the naive quark model. To understand the nature of these
scalars has been the focus of recent studies \textit{e.g.} see Refs. \cite{pdg2010,Scalar,Scalar-lightGB,Glueball-Review}
and references therein.

Among the low-lying scalar mesons, the lightest scalar $f_0(600)$ or $\sigma$ attracts a lot of
interests. It is widely believed that $f_0(600)$ is like the Higgs boson
which plays a crucial role in the spontaneous chiral symmetry breaking. Confirmation
of existence of the elusive $f_0(600)$ from $\pi \pi$ scattering processes 
settles down a controversy last for more than a few decades \cite{Caprini:2005zr,Scalar}. 
The $\pi K$ scattering \cite{Black:1998wt} and analysis from 
D decay $D^+ \to K^- \pi^+ \pi^- $ \cite{Bugg:2009uk} revealed that $\kappa$ 
should also exist. BES II also found such a $\kappa$ like structure in $J/\Psi$ decays \cite{Ablikim:2005ni}. 
Combined with the well determined sharp resonances, 
i.e. isoscalar $f^0(980)$ and isotriplet $a(980)$ from $\pi \pi$ and $\pi \eta$ as 
well as $K K$ scattering processes, now it is accepted in literature that
these low-lying scalar mesons (say less than $1\,\,\textrm{GeV}$) can be cast into a chiral nonet. 
The next important issue is what is the nature of this nonet.

There are a couple of viewpoints on the nature of
this nonet. For example,the tetraquark model \cite{jaffe} can explain the
mass hierarchy and decay pattern of this nonet quite successfully and is further supported from
other experimental data, like the photon-photon collision data,
which prefer the tetraquark interpretation for the lowest scalar
meson nonet \cite{Achasov:2009ee} (where it is demonstrated that
$f_0(980)$ should be a tetraquark dominant state with great
details). An alternative interpretation is that this nonet is bound state of the meson-meson
molecule \cite{molecule}. In any way, this nonet challenges a self-consistent interpretation in the naive quark model.

Nonetheless, the agreement on the nature of this nonet has not
been achieved yet. For example, recently by studying $\pi\pi$ and
$\gamma \gamma$ scatterings, it is found that this particle could
have a sizable fraction of glueball  \cite{Minkowski:1998mf,arXiv:0804.4452}. The
K-matrix analysis \cite{Mennessier:2010xg} suggested that
$f_0(600)/\sigma$ should be a glueball dominant state while
$f_0(980)$ should be a mixture of tetraquark and $q{\bar q}$. A
recent pole analysis with Pad{\'e} approximation suggests $f_0(980)$
might be more like a molecular state \cite{Dai:2011bs}. 
The nature of this nonet is also a focus of lattice study \cite{Alford:2000mm}.
For example, the physical state of $\sigma$ and $\kappa$ can have sizable 
tetraquark components, as demonstrated in a recent lattice simulation \cite{Prelovsek:2010kg}.

The great success of chiral symmetry in understanding the nature of
lightest pseudo-scalars motivates us to the extended the linear $\sigma$ models \cite{qishu,Fariborz:2011es}
to study the nature of these low-lying isoscalar and pseudoscalar (say, scalars less than $2$ GeV. The historic review on scalars above 1 GeV but below 2 GeV can be found in \cite{pdg2010}). Such models may 
shed some lights on the nature of light isoscalars, especially on the issue of mixing 
among glueball, quarkonia and tetraquark. One interesting question which can be addressed
by such models is that which of these low-lying iso-scalar and pseudoscalar states 
are more glueball-like. The results from 
lattice simulations suggest $f_0(1500)$ or $f_0(1700)$ could be a glueball 
rich iso-scalar while $\eta(1489)$ can be a glueball rich
pseudoscalar \cite{Mathieu:2008me}. Since the $0^{++}$ and $0^{-+}$ 
glueball states can significantly mix with
quarkonia and tetraquark states of the same quantum numbers, 
it is necessary to include these states and a glueball state in an extended linear $\sigma$ model.

In this work, we extend our previous work \cite{qishu} by 
including a nonet to accommodate the tetraquark states and 
focusing on the mixing of quarkonia and tetraquark. 
We attempt to address on the issues which states in the 
pseudoscalar and isoscalar sectors are most glueball-like. 
Comparing with the systematic work shown in \cite{Fariborz:2011es} 
and the references therein, we extend the linear $\sigma$ model by 
including a complex singlet field as a pure glueball field and 
introduce a determinant interaction term \cite{u11}
instead of the logarithmic term to solve the $U(1)_A$ problem as in \cite{fariborz1}. 
Our model predicts that the isoscalar glueball should be heavier than $2.0$ GeV
when the pseudoscalar $\eta(1726)$ is the best glueball candidates. 
The lowest isoscalar $f_0(600)$ is found to be quarkonia dominant state with 
a considerable tetraquark component.

The rest of the paper is organized as follows. In section 2, we introduce the extended linear $\sigma$ model and
derive the vacuum conditions for the condensates of quarkonia and glueball fields. In section 3 we present our 
detailed numerical analysis. In section 4 we close our study with discussions and conclusions. An appendix is provided
to show the mass matrices of isotriplet, isodoublet and isosinglet states.

\section{The extended linear $\sigma$ model}

The extended linear $\sigma$ model can be systematically formulated under the
symmetry group $SU_R(3) \times SU_L(3) \times U_A(1)$. 
Three types of chiral fields are included: a $3\times3$
matrix field $\Phi$ which denotes the quarkonia states, a $3\times3$
matrix field $\Phi^\prime$ which denotes the tetraquark state, and a
complex field $Y$ which denotes the pure glue-ball states and is 
a chiral singlet. The transformation properties of these fields under the chiral symmetry
are defined as follows:

\begin{align}
\Phi & \rightarrow U_L \Phi {U^\dagger}_R, \nonumber \\
\Phi^\prime &\rightarrow  U_L \Phi^\prime {U^\dagger}_R\,\,,
\end{align}
where $U_{L,R}$ are group elements of the $SU(3)_L \times SU(3)_R$ symmetry. 
While the complex field $Y$ is invariant under this $SU_R(3) \times SU_L(3)$ transformation. 
Under the $U_A(1)$ transformation, each
fields is changed by a global phase factor  as defined below
\begin{align}
\Phi &\rightarrow e^{2 i \theta_A} \Phi,  \nonumber \\
\Phi^\prime &\rightarrow e^{-4 i \theta_A} \Phi^\prime \,\,, \nonumber \\
Y & \rightarrow e^{- 6 i \theta_A} Y\,\,.
\end{align}

Following the convention of the linear sigma model, we express the quarkonia fields,
the tetraquark fields, and the glueball fields as:
\begin{align}
\Phi &= T_a \phi_a = T_a (\sigma_a + i \pi_a), \notag \\
\Phi^\prime &= T_a {\phi^\prime}_a = T_a ({ \sigma^\prime}_a+ i {\pi^\prime}_a),
\notag \\
Y &=\frac{1}{\sqrt  2} (y_1 + i\, y_2 ) \,,
\end{align}
where matrices $T_a = \frac{\Lambda_a}{2}$ are the generators of
$U(3)$ and $\Lambda_a$ are the Gell-Mann matrices with $\Lambda_0 =
\sqrt \frac{2}{3} \,\,1_{3\times3}$. Fields, $\sigma_a$ and $\pi_a$,
$\sigma_a^\prime$ and $\pi^\prime_a$, and $y$ and $a$, denote
quarkonia, tetraquark and glueball states in the chiral basis,
respectively.

Up to the mass dimension $O(p^4)$ (it is
believed they are the most important operators to determine the
nature of light scalars of ground states), the Lagrangian of our model can
include two parts: the symmetry invariant part $\cal{L}_S$
and the symmetry breaking one $\cal{L}_{SB}$:
\begin{align}
\cal{L} &= \cal{L}_S + \cal{L}_{SB},
\label{totsl}
\end{align}
where the symmetry invariant part includes the following terms
\begin{align}
\label{sl} \cal{L}_{S} &= Tr(\partial_\mu \Phi \partial^{\mu}
\Phi^{\dagger})+ Tr(\partial_\mu \Phi^\prime \partial^{\mu}
\Phi^{\dagger \prime})+
\partial_\mu Y \partial^\mu Y^\star
 \notag  \\
 & -{m_\Phi}^2 Tr (\Phi^\dagger \Phi)
-{m_{\Phi^\prime}}^2 Tr (\Phi^{\dagger \prime} \Phi^\prime) -{m_Y}^2 Y Y^\star
\notag \\
& -\lambda_1 Tr (\Phi^\dagger \Phi \Phi^\dagger \Phi)
 -{\lambda_1}^\prime
Tr (\Phi^{\dagger \prime} \Phi^\prime \Phi^{\dagger \prime} \Phi^\prime)
-\lambda_2 Tr (\Phi^\dagger \Phi \Phi^{\dagger \prime} \Phi^\prime)
- \lambda_Y (Y Y^\star)^2
\notag \\
& - [ \lambda_3 \epsilon_{abc} \epsilon^{def} {\Phi_d}^a {\Phi_e}^b
{\Phi_f}^{\prime c} + h.c.] + [ k Y Det(\Phi) + h.c.],
\end{align}
while the symmetry breaking part includes the following terms
\begin{align}
\cal{L}_{SB} &= [Tr(B.\Phi) + h.c.] + [Tr(B^\prime.\Phi^\prime) +
h.c.]  + (D.Y +h.c.) - [ \lambda_m Tr(\Phi \Phi^{\dagger \prime})
+h.c. ].
\label{sbt}
\end{align}
As evident from our Lagrangian, the choice of these operators are
not an exhaustive one and many more terms are allowed, as demonstrated in
\cite{Fariborz:2011es}. In spite of that,
these terms can be considered as a leading choice when considering 
the number of quark plus antiquark lines at an effective
vertex as argued in \cite{fariborz1}, where the authors had
restricted the maximum number of quark plus antiquark lines up to 8.
Here we relax such a constraint by including two terms
related with the tetraquark fields. 

At this point we would like make some comments on the interaction or
mixing terms in our Lagrangian. We have considered direct mixing
terms between quarkonia with tetraquark fields and an interaction term between the quarkonia
with glueball fields while neglected the direct coupling between the
tetraquark field and the glueball.

The mixing between quarkonia and tetraquark is affected by three terms in
this work which are quadratic (proportional to the coupling $\lambda_m$), cubic (proportional to the coupling $\lambda_3$) and quartic (proportional to the coupling $\lambda_2$) in effective fields.
The quadratic interaction term in Eq. (\ref{sbt}) can be considered as an effective mixed mass term
which violated the $U(1)_A$ symmetry, which is a
higher order term if we count in terms of number of internal quark plus
anti-quark lines. But we do not differentiate between
quarkonia and tetraquark effective fields on their underlying quark content and
treat them on the same footing as effective fields. So we retained these two higher order 
terms in our choice of Lagrangian. Among these three mixing terms
only the cubic interaction term has been considered in the reference \cite{fariborz1}.

The other term which is different from the work of \cite{fariborz1}
is the last term (proportional to the coupling $\kappa$) in Eq. (\ref{sl}). This choice is motivated from the 
observation of 't Hooft, who introduced the coupling between a scalar spurion field to the
determinant of the quarkonia field in order to solve the $U(1)_A$ problem. Furthermore,
the study of Lattice QCD and sum rules reveals that the instanton effect plays an important 
role in shaping the properties of the glueball ground state, it is 
well-justified to consider this spurion field as an effective glueball field, as argued in \cite{qishu}.

It is worth mentioning here that we shall also consider the mixing terms like $Tr[\Phi^\dagger \Phi]Y^* Y$ and
$Y Tr(\Phi {\Phi^{\dagger \prime}})$. For simplicity, we drop those terms in this work but will be studied in future. 

Except the explicit symmetry breaking terms, the chiral symmetry and $U_A(1)$ symmetry 
are further broken through the formation of condensates.
Both quarkonic and tetraquarkonic condensates are responsible for the
spontaneous breaking of the chiral symmetry from $SU_L(3)
\times SU_R(3)$ to $SU_V(2)$ whereas gluonic
condensate for the $U(1)_A$ symmetry. The remnant
isospin allows us to represent two condensates for quarkonia and
teraquark fields each as: $v_0$, $v_8$ and ${v^\prime}_0$,
${v^\prime}_8$. While the gluonic condensate in our theory is 
labelled as $v_y$.

Expanding fields around these vacuum expectation values we get the expression
of tree level potential $V(v_0, v_8, {v^\prime}_0, {v^\prime}_8,
v_y)$ which should be stable under the variation of condensates,
i.e.,
\begin{align}
\frac{\partial V(v_i, {v^\prime}_i, v_y )}
{\partial (v_i,{v^\prime}_i,v_y)} = 0, &&  i=0,8 \label{vac.stable}
\end{align}

The explicit expressions for each equations can be worked out straightforwardly and are given below:
\begin{align}
\frac{\partial V}{\partial v_0} &= b_0 + \frac{1}{4\sqrt{3}}( 2 {v_0}^2 -
{v_8}^2)v_y k -v_0 {m_\Phi}^2 -(\frac{{v_0}^3}{3} + v_0 {v_8}^2
-\frac{{v_8}^3}{3\sqrt{2}} )\lambda_1  \notag \\
&
- \frac{1}{3} (v_8 {v_0}^\prime {v_8}^\prime - \frac{v_8 {v_8}^{ \prime 2}}
{2 \sqrt{2}} + \frac{v_0 {v_0}^{\prime 2}}{2}
+\frac{v_0 {v_8}^{\prime 2}}{2} )\lambda_2 -\sqrt{\frac{2}{3}}
(2 v_0 {v_0}^\prime -v_8 {v_8}^\prime ) \lambda_3 -\frac{{v_0}^\prime}{2}
\lambda_m \label{vc1}
\end{align} \begin{align}
\frac{\partial V}{\partial v_8} &= b_8 -\frac{1}{2\sqrt{3}} v_8 ( v_0 
+\frac{v_8}{\sqrt{2}})v_y k - v_8 {m_\Phi}^2 -({v_0}^2 v_8 - \frac{v_0 {v_8}^2}{\sqrt{2}}
+ \frac{ {v_8}^3}{2} )\lambda_1 \notag \\ & -( \frac{v_8 {v_0}^{\prime 2}}{6}
- \frac{v_8 {v_0}^\prime {v_8}^\prime }{3 \sqrt{2}}
 + \frac{v_8 {v_8}^{\prime 2}}{4} + \frac{v_0 {v_0}^\prime {v_8}^\prime}{3}
 -\frac{v_0 {v_8}^{\prime 2}}{6 \sqrt{2}} )\lambda_2
+ \sqrt{\frac{2}{3}} (v_8 {v_0}^\prime + v_0 {v_8}^\prime + \sqrt{2}
v_8 {v_8}^\prime) \lambda_3 - \frac{{v_8}^\prime}{2} \lambda_m  
 \end{align}\begin{align}
\frac{\partial V}{\partial {v_0}^\prime} &= {b_0}^\prime -{v_0}^\prime
{m_{\Phi^\prime}}^2 - \frac{1}{3} (\frac{{v_0}^2 {v_0}^\prime }{2}
+ \frac{{v_8}^2 {v_0}^\prime }{2} +v_0 v_8 {v_8}^\prime
-\frac{ {v_8}^ 2 {v_8}^\prime}{2 \sqrt{2}} ) \lambda_2
 \notag \\& - \frac{1}{\sqrt{3}} (\sqrt{2} {v_0}^2
- \frac{{v_8}^2}{\sqrt{2}} ) \lambda_3
- (\frac{{v_0}^{\prime 3}}{3} + {v_0}^\prime {v_8}^{\prime 2}
-\frac{{v_8}^{\prime 3}}{3 \sqrt{2}} ){\lambda_1}^\prime
 -\frac{v_0}{2} \lambda_m 
 \end{align}\begin{align}
\frac{\partial V}{\partial {v_8}^\prime} &= {b_8}^\prime -{v_8}^\prime
{m_{\Phi^\prime}}^2 + (\frac{{v_8}^2 {v_0}^\prime }{6 \sqrt{2}}
- \frac{{v_0}^2 {v_8}^\prime }{6}-\frac{{v_8}^2 {v_8}^\prime}{4}
-\frac{v_0 v_8 {v_0}^\prime}{3} +\frac{ v_0 v_8 {v_8}^\prime}
{3 \sqrt{2}} ) \lambda_2
 \notag \\ & + \frac{1}{\sqrt{3}} (\sqrt{2} v_0 v_8 + {v_8}^2 ) \lambda_3
- ({v_0}^{\prime 2} - \frac{ {v_0}^\prime {v_8}^{\prime} }{\sqrt{2}}
+\frac{{v_8}^{\prime 2}}{2} ){v_8}^\prime {\lambda_1}^\prime
 -\frac{v_8}{2} \lambda_m 
 \end{align}\begin{align} \frac{\partial V}{\partial v_y} &= \sqrt{2}D + \frac{1}{2 \sqrt{3}}
(\frac{ {v_0}^3}{3} -\frac{v_0 {v_8}^2}{2} -\frac{{v_8}^3}{3 \sqrt{2}})k
-v_y {m_y}^2 -{v_y}^3 \lambda_Y \label{vc5}
\end{align}

These five constraints are nonlinear in terms of condensates $v_0$,
$v_8$, and $v_0^\prime$, $v_8^\prime$, $v_y$, but are linear in
terms of couplings. To avoid solving nonlinear equations, in our numerical analysis, we
can choose a set of $v_0$, $v_8$, and $v_0^\prime$, $v_8^\prime$, $v_y$
as input to solve couplings.

More precisely, in order to guarantee that our values of $\{v_0^\prime,v_8^\prime\}$ are physically meaningful,
we replace them by two positive quantities, i.e. \{$v_q^\prime$,$v_s^\prime$\}. The relation between $\{v_0^\prime
,v_8^\prime\}$ and $\{v_q^\prime,v_s^\prime\}$ can be found from
\cite{Schaefer:2008hk}, and are provided below as
\begin{align}
v_0^\prime = \frac{\sqrt{2}}{\sqrt{3}} v_q^\prime + \frac{1}{\sqrt{3}} v_s^\prime\, \,,v_8^\prime = \frac{1}{\sqrt{3}} v_q^\prime - \frac{\sqrt{2}}{\sqrt{3}} v_s^\prime\,.
\end{align}
The quarkonia condenstates $v_0$ and $v_8$ are solved out from the decay constants of Pion and Kaon, which are given below
\begin{align}
f_\pi =(\frac{\sqrt{2}}{\sqrt{3}} v_0+ \frac{1}{\sqrt{3}} v_8) \cos \theta_\pi - (\frac{\sqrt{2}}{\sqrt{3}} v_0^\prime + \frac{1}{\sqrt{3}} v_8^\prime)  \sin \theta_\pi \,,\\
f_K =(\frac{\sqrt{2}}{\sqrt{3}} v_0- \frac{1}{\sqrt{12}} v_8) \cos \theta_\pi - (\frac{\sqrt{2}}{\sqrt{3}} v_0^\prime - \frac{1}{\sqrt{12}} v_8^\prime)  \sin \theta_\pi\,.
 \end{align}
These two relations between the decay constants and our model parameters can be found by constructing the Noether current and utilizing the PACA relations, as demonstrated in \cite{Lenaghan:2000ey}.

\section{Numerical Analysis}

Due to the unbroken $SU(2)_V$ isospin symmetry, physical scalar
and pseudo-scalar states can be categorized into three groups with isospin quantum
numbers as $I = 1$ (triplet), $\frac{1}{2}$ (doublet) and $0$, respectively.
Only bare quarkonia, tetraquark and glueball fields with the same
isospin quantum number can mix with each other to form physical
states. Moreover, there is no mixing between scalar and pseudoscalar
fields. Thus the chiral singlet glueball field can only mix with the
isospin singlets of quarkonia and tetraquark fields. 

Using these facts, the physics states below 2 GeV can be tabulated as given in Table \ref{mixing}, 
\begin{table}
\begin{tabular}{ |c|c|c|c| }
\hline
Isospin & $I=1$ & $I=\frac{1}{2}$ & $I=0$  \\ \hline
PseudoScalars(P=-1)& $\{\pi, \pi^\prime \}$ & \{K, $K^\prime$\}, \{$K^*, K^{*\prime} $\} & \{$\eta_1, \eta_2, \eta_3,
\eta_4, \eta_5$ \} \\ \hline
Scalars(P=1)& \{$a$, $a^\prime$ \}, & \{$\kappa$, $\kappa^\prime$\}, \{$\kappa^*$, $\kappa^{*\prime}$\}, & $\{f_1,f_2,f_3,f_4,f_5\}$ \\ \hline
\end{tabular}
\caption{The categorization of scalar and pseudo-scalar states in term of isospin quantum number are demonstrated. 
States in the same category can mix with each other.}
\label{mixing}
\end{table}
where the isodoublet $\{K, K^\prime\}$ is connected with the isodoublet \{$K^*,
K^{*\prime} $\} by charge conjugation. And a similar relation holds
for $\{\kappa$, $\kappa^\prime\}$ and $\{\kappa^*$,
$\kappa^{*\prime}\}$.

For both isotriplet and isodoublet sector, a $2 \times 2$ mixing
matrix can be used to describe the mixing among quarkonia and tetraquark states. While for the isosinglet scalar and
pesudo-scalar sectors,  a $5 \times 5$ mixing matrix must be used to describe such a mixing. 
To extract the relevant mass matrices, we use the following substitutions $\sigma_0
\to v_0 + \sigma_0$, $\sigma_8 \to v_8 + \sigma_8$, $\sigma_0^\prime
\to v_0^\prime + \sigma_0^\prime$, $\sigma_8^\prime \to v_8^\prime +
\sigma_8^\prime$, and $y_1 \to v_y + y_1$ while assuming that other
fields have no vacuum expectation value. These mass matrices are
provided in the Appendix.

It is useful to count the total number of free parameters in our model.
There are 15 free parameters as shown in our extended linear $\sigma$ model
given in Eq. (\ref{totsl}). Most of these parameters can
be fixed by the input from the isotriplet and isodoublet sectors. The only unfixed 
parameters are related to glueball sector. Meanwhile, five vacuum
stability conditions further reduce the number of free parameters.
Therefore our model is over-constrained by experimental data 
except the glueball sector. Below
we describe how to fix our model parameters.

\begin{itemize}
\item The tetraquark vacuum condensates $\{v_q^\prime\,,v_s^\prime\}$
are treated as input and are assumed to be positive but smaller than 2 GeV. 

\item To fix the parameters in the triplet and doublet sector, we use
the physical masses and decay constants of $\{\pi, \pi^\prime\}$ and $\{K, K^\prime\}$ mesons,
as shown in Table \ref{triplet}.
\begin{table}
\begin{tabular}{ |c|c|c|c|c|c|c|}
\hline
Fields  &$\pi$ & $\pi^\prime$  & $f_\pi$  & K & $K^\prime$& $f_K$ \\\hline
Mass (GeV) & 0.14 & 1.20-1.40  & 0.15 & 0.49 & 1.46  & 0.13\\ \hline
\end{tabular}
\caption{The experimental masses and decay constants of a triplet and a doublet are provided.}
\label{triplet}
\end{table}
The mixing angles for isotriplets and isodoublets, which are labelled as $\theta_{K}$ and $\theta_{\pi}$,
are treated as input and are restricted to vary in the range $\{
-\frac{\pi}{4}, \frac{\pi}{4} \}$ (since it is widely believed that Pion and Kaon are quarkonia states. We will address this issue in our discussions). Accordingly, four free parameters in our model are
fixed with another two free parameters are traded off by two mixing angles. We also impose the constraints on the trace of 
mass matrix of isotriplets $a$ and $a^\prime$, i.e. $Tr[M_a^2]=\sum (M_a^2)^{Exp}$ and the trace of mass matrix of isodoublet $\kappa$ and $\kappa^\prime$, i.e. $Tr[M_\kappa^2]=\sum (M_\kappa^2)^{Exp}$. With these constraints, 
we choose parameters $\{v_0,v_8, m_\Phi^2, (m_\Phi^\prime)^2, \lambda_1, \lambda_2, \lambda_3,\lambda_1^\prime, \lambda_m, k \,v_y \}$ as solved out from input of isotriplet and isodoublet sectors.

\item In order to further constrain the parameters related to glueball sector,
following the method in the reference \cite{fariborz2}, we consider
two broad conditions from the isoscalar pseudo-scalar sector
\begin{align}
Tr[{M_\eta}^2]_{Model} = Tr[{M_\eta}^2]_{Exp}\,\,, \label{trace} \\
Det[{M_\eta}^2]_{Model} = Det[{M_\eta}^2]_{Exp}\,\,.
\end{align}
where, $M_\eta$ is the mass matrix for the isoscalar pseudo-scalar states. So
two more free parameters are fixed and we choose $\{k, m_Y^2 + \lambda_Y v_Y^2\}$ as solved from these two constraints. Combined with the solution of $k \,v_Y$ given in the previous step, the parameter $v_Y$ is solved out.

\item The five vacuum stability conditions given in Eqs. (\ref{vc1}-\ref{vc5}) can further help to reduce free parameters in our model. In practice, we choose the following five free parameters $\{b_0,b_8,b_0^\prime,b_8^\prime, D\}$ as solved from these five equations.

\end{itemize}


After these input, there is one free parameter unfixed which is selected as 
the glueball mass $m_Y^2$. By using it as input, we can predict
the masses of lowest scalar and pseudoscalar as well as their
components. 

\begin{table}
\begin{tabular}{ |c|c|c|c|c|c|c|c|c|c|c|c|c|c|c|}
\hline
Fields& a & $a^\prime$ & $\kappa$& $\kappa^\prime$ & $\eta_1$ & $\eta_2$ &
$\eta_3$ & $\eta_4$ & $\eta_5$  & $f^0_1$ & $f^0_2$ & $f^0_3$ & $f^0_4$& $f^0_5$\\ \hline
Mass(GeV)& 0.98& 1.47 & 0.80 & 1.43 & 0.55 & 0.96 & 1.30  & 1.48 & 1.76 & 0.4-1.2& 0.98& 1.2-1.5 & 1.505&  1.72\\ \hline
\end{tabular}
\caption{The experimental mass spectra for triplets, doublets and iso-scalars are used to determine the best fit.}
\label{expinput}
\end{table}


Considering that there is a large uncertainty in the determination of $\pi^\prime$ mass, we
choose to vary this mass within the range $1.2-1.4 \text{GeV}$. We scan $5$ mass values in 
a $50$ $\text{MeV}$ step within the above specified range starting with $1.2$ $\text{GeV}$.

To test our model, here we are going to
demonstrate, how well it can reproduce the mass spectra
of mesons. Besides that we are also interested to see
what is the composition of these low-lying scalar. 

There is a huge number of possible solutions in our parameter space.
We regard that the best solution for the parameter set is the one which
closely reproduces the mass spectra of scalars close to the
experimental measured values. The goodness of the solution is on the basis of
smallness of the below defined two quantities: 
the first one is $\chi_1$ as defined in \cite{fariborz2} 
\begin{align}
\chi_1 = \displaystyle\sum\limits_{i=1}^{13} \frac{\Bigl|{M_i}^{theo} -
{M_i}^{exp}\Bigr|}{{M_i}^{exp}}, 
\label{chi1}
\end{align}
and we also consider the
second one which is defined by the least $\chi^2$ method labelled as $\chi_2$ below:
\begin{align}
\chi_2 = \displaystyle\sum\limits_{i=1}^{13} \frac{\Bigl|{M_i}^{theo} -
{M_i}^{exp}\Bigr|^2}{(\delta {M_i}^{exp})^2},
\label{chi2}
\end{align}
where, ${M_i}^{theo(exp)}$ is the mass of the each member of scalar
or pesudo-scalar family calculated from our model (experiment).
whereas, $\delta {M_i}^{exp}$ is the experimental error for each
mass. The sum takes into account $5$ pseudo scalar masses, 4 scalar masses,
the masses of two triplets $a$ and $a^\prime$, and the masses of two doublets $\kappa$ and $\kappa^\prime$.

We also take into account the decay width of the lowest scalar
$f_0(600)\rightarrow \pi \pi$ as a constraint. This decay width at the
tree level are computed and those solutions which give the decay width between
$0.35-0.9 \text{GeV}$ are regarded as reasonable.

Below we explain how we choose our best solution. For this
we apply two types of minimum $\chi_1$ and $\chi_2$ analysis in
two stages.Typically $\chi_1$ has a smaller value while $\chi_2$ has a larger value due to the small experimental errors
for some scalars, like the $\eta_1$ and $\eta_2$.
Firstly, for a given mass value of $m_{\pi^\prime}$ we choose the best fit solution for the mass spectra for all 
scalars. We generate more than $8$ million random parameter sets of $\{\theta_\pi, \theta_K, v_q^\prime, v_s^\prime\}$ to find the best fit solution which minimizes the $\chi_i$.
For each value of $m_\pi{^\prime}$ and each set of $\{\theta_\pi, \theta_K, v_q^\prime, v_s^\prime\}$, we treat the bare glueball mass $m_Y^2$ a scanning parameter varying from $-9$ to $9$. After considering the constraint the vacuum must be bounded below, i.e. the coupling $\lambda_Y$ must be positive, we read out the best fit value for $m_Y^2$. 
Then we determine the best fit for each scanned $m_{\pi^\prime}$ from $1.2$ GeV to $1.4$  GeV with $50$ MeV interval. The final best fit is chosen from these 5 best fits with the minimum $\chi_1$/$\chi_2$ value.



The different $\chi_i$ definitions given in Eqs. (\ref{chi1}-\ref{chi2}) yield a similar bestfit, which are presented in Table (\ref{chi1-prm}). We would like to highlight a few features
out from it. 1) It is  the negative mass parameter $m_{\Phi}^2$ that triggers the chiral symmetry breaking. 2) The sign of $v_Y$ is correlated with the sign of $k$, and the sign of $k$ is determined from the mass spectra of pseudoscalar sector. 3) The couplings $\lambda_1$, $\lambda_1^\prime$, $\lambda_Y$ are positive which guarantee the potential is bounded from below. 4) The values of $\lambda_1$, $\lambda_2$, and $\lambda_Y$ as well as $k$ are large, which demonstrate the non-perturbation nature of the model. 5) The value of $\lambda_m$ is found to be negative. 
\begin{table}[h]
\begin{tabular}{ |c|c||c|c| }  
\hline
Parameter & Value & Parameter & Value  \\ \hline
$\theta_{\pi}$ (radian) & -0.604   & ${\lambda_1}^\prime$     & 8.248    \\  \hline
$\theta_{K}$ (radian)     & -0.714   & $\lambda_2$              & 76.428    \\ \hline
$v_0$ (GeV)                          & 0.074    & $\lambda_3$ (GeV)        & -0.738   \\ \hline
$v_8$ (GeV)                          & -0.115   & $\lambda_Y$              & 38.327    \\ \hline
${v_0}^{\prime}$ (GeV)               & 0.203    & k                        & -78.15   \\ \hline
${v_8}^{\prime}$ (GeV)               & 0.126    & $\lambda_m$ ($GeV^2$)    & -1.044   \\ \hline
$v_y$ (GeV)                          & -0.109   & $b_0$ ($GeV^3$)          & -0.085    \\ \hline
${m_Y}^2$ ($GeV^2$)                  & 3.0      & $b_8$ ($GeV^3$)          & -0.161   \\ \hline
${m_{\Phi}}^2$ ($GeV^2$)             & -0.025   & ${b_0}^\prime$ ($GeV^3$) & 0.166    \\ \hline
${m_{\Phi^\prime}}^2$ ($GeV^2$)      & 0.744    & ${b_8}^\prime$ ($GeV^3$) & 0.18   \\ \hline
$\lambda_1$                          & 35.465   & D ($GeV^3$)              & -0.265   \\ \hline
\end{tabular}
\caption{The values of parameters in our fit are shown where the best value of $m_{\pi{^\prime}}$ is found to be $m_{\pi^\prime} = 1.2$ GeV.}
\label{chi1-prm}   
\end{table}

Our best fit result favor the case where the percentage of tetra quark component in $\pi^\prime$ meson is about $67.7\%$ and in $K^\prime$ meson is about $57.2\%$.  When comparing our result with
the previous studies, we find that the tetraquark component of
$K^\prime(1.46)$ in our result is quite low compared to $95\%$ in
\cite{giacosa} and $76\%$ in \cite{fariborz2}. It would be
attributed to effects of glueball (note: though glueball does not directly
mixed with the other fields in doublet sector, the parameters like
gluonic condensate and the instanton coupling constant do contribute
to the mass matrix in the doublet sector, as shown up by $k \, v_Y$) or decay widths of these
mesons will put a more stringent constraint on this percentage. For
$\pi^\prime$ the percentage of the tetraquark component is in
qualitative agreement with that of \cite{fariborz2} where they found
a tetraquark percentage of $85\%$.

With the parameter set given in Table (\ref{chi1-prm}), the predicted mass spectra and components for the triplet $\{a,a^\prime \}$ and the doublet $\{\kappa, \kappa^\prime \}$ are provided in Table \ref{chi1-doub}. It is found that the mass spectra of the triplet are close to their experimental values but those of the doublet deviate considerably. The $a$ and $\kappa^\prime$ are more tetraquark-like while $a^\prime$ and $\kappa$ are more quarkonia like.

\begin{table}[h]
\begin{tabular}[b]{ |c|c|c|c|c|c| }
  \hline
$\pi^\prime$ Mass (GeV) & Field & Our Value (GeV) & quarkonia ($\%$) & tetraquark ($\%$)& Experimental Value (GeV)  \\ \hline
     & $a$                & 1.055  & 38.14   & 61.8 6 & 0.98 \\  \cline{2-6}
     & $a^{\prime}$       & 1.417  & 61.86   & 38.14 & 1.47 \\ \cline{2-6}
1.2  & $\kappa$           & 1.13   & 62.14   & 37.86 & 0.80 \\ \cline{2-6}
     & $\kappa^{\prime}$  & 1.186   & 37.86   & 62.14& 1.43 \\ \hline
\end{tabular}
\caption{Mass spectra and components for the triplet and doublet sector based on our fit are demonstrated where the best value of $m_{\pi{^\prime}}$ is found to be $m_{\pi^\prime} = 1.2$ GeV.}
\label{chi1-doub}
\end{table}

\begin{figure}[t]
\centerline{
\epsfxsize=7 cm \epsfysize=5 cm \epsfbox{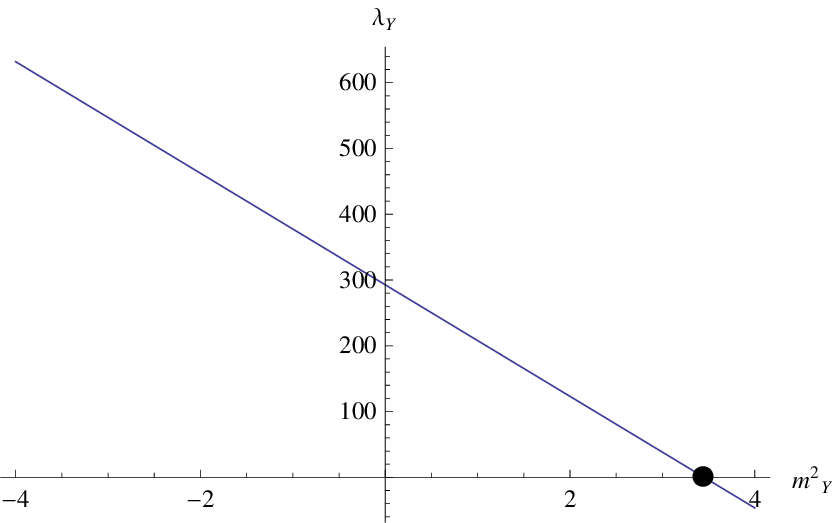}
\hspace*{0.2cm}
\epsfxsize=7 cm \epsfysize=5 cm \epsfbox{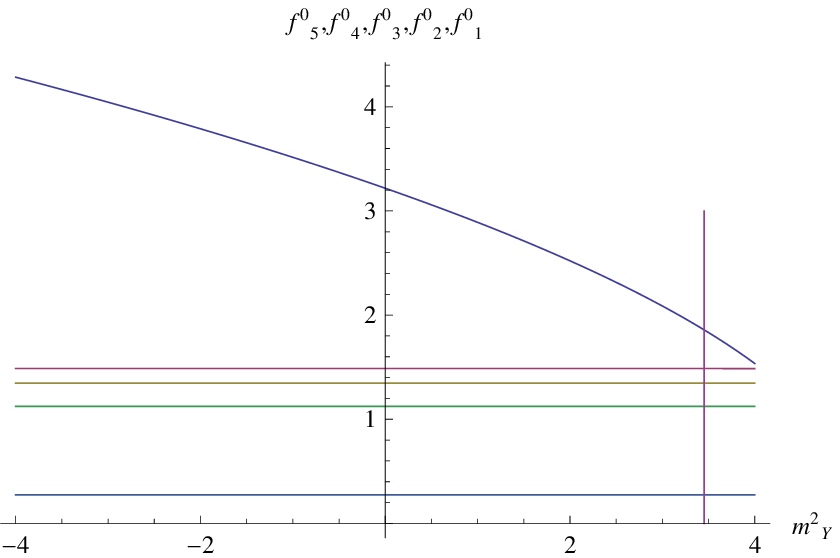}}
\vskip -0.05cm \hskip -0.07 cm \textbf{( a ) } \hskip 6.7 cm \textbf{( b )}
\caption{a) The dependence of $\lambda_Y$ upon $m_Y^2$ is demonstrated. A solid circle marker shows the point $\lambda_Y=0$, which corresponds to $m_Y^2=3.452$ ($m_Y=1.858$) . b) The dependence of mass of $f_0$ upon $m_Y^2$ is demonstrated. A vertical line with $m_Y^2$ is drawn to read out the lowest mass $m_{f_0^5}=1.86$.}
\label{figf}
\end{figure}

The predicted mass spectra of pseudoscalars and scalars are shown in Tables (\ref{chi1-psc}-\ref{chi1-sc}). We have a few comments in order. 1) The pseudo-scalar mass spectra can fit experimental data better than the scalar mass spectra but the mixing pattern of scalar and pseudo-scalar are quite similar. 2) When the mass $m_{f_5^0}=1.72$ GeV is used to our fit, we find a solution with $\lambda_Y<0$. To guarantee the condition that the potential must be bound from below (i.e. $\lambda_Y>0$), we have to keep $m_{f^0_5}$ out from our fit, which explains why in the definition of $\chi_i$ we only sum over the masses of 4 scalar. It is found that this condition can predict the lightest glueball scalar should be around 2.0 GeV or so, as can be read off from Fig. (\ref{figf}b), while the lightest glueball pseudo scalar should be $\eta_1$. The mass splitting between these two glueball states is controlled by parameters $v_Y$ and $\lambda_Y$ and is found to be around $0.15$ GeV. When compared with the Lattice QCD prediction for the glueball bare mass reported in \cite{lattice} where the mass is $1.611$ GeV, our result $m_Y=1.73$ GeV is slightly heavier than this prediction. When $m_Y=1.611$ GeV is taken, then the predicted mass of the lightest glueball is $m_{f^5_0}=2.29$ GeV. 3) The lightest scalar $f^0(600)$ is found to be $0.27$ GeV or so and is a quarkonia dominant state. 
\begin{table}[h]
\begin{tabular}{ |c|c|c|c|c|c|c| }   
\hline
$\pi^\prime$ Mass (GeV) & $J^{PC} = 0^{-+}$ & Our Value (GeV)  & quarkonia ($\%$) & tetraquark ($\%$) & glueball ($\%$)  & Experimental Value (GeV) \\ \hline
     & $\eta_5$ & 1.858  & 0.037 & 0.001 & 99.962 & 1.756 $\pm$ 0.009 \\  \cline{2-7}
     & $\eta_4$ & 1.380  & 75.803 & 24.167 & 0.03  & 1.476 $\pm$ 0.004 \\ \cline{2-7}
1.2  & $\eta_3$ & 1.291  & 26.700 & 73.294 & 0.006 & 1.294 $\pm$ 0.004 \\ \cline{2-7}
     & $\eta_2$ & 0.907   & 15.852 & 84.145 & 0.003  & 0.95766 $\pm$ 0.00024 \\ \cline{2-7}
     & $\eta_1$ & 0.595   & 81.607 & 18.393 & 0.0 & 0.547853 $\pm$ 0.000024 \\ \hline
\end{tabular}
\caption{Mass spectra and components for the pseudo-scalar mesons based on our fit are shown where the best value of $m_{\pi{^\prime}}$ is found to be $m_{\pi^\prime} = 1.2$ GeV.}
\label{chi1-psc}
\end{table}

\begin{table}[h]                 
\begin{tabular}{ |c|c|c|c|c|c|c| }     
  \hline
$\pi^\prime$ Mass (GeV) & $J^{PC} = 0^{++}$ & Our Value (GeV) & quarkonia ($\%$) & tetraquark ($\%$) & glueball ($\%$) & Experimental Value (GeV) \\ \hline
     & ${f_5}^0$ & 2.09   & 0.01 & 0.0 & 99.99  & - \\  \cline{2-7}
     & ${f_4}^0$ & 1.487   & 77.469 & 22.53 & 0.001 & 1.505 $\pm$ 0.006 \\ \cline{2-7}
1.2  & ${f_3}^0$ & 1.347  & 22.177 & 77.82 & 0.003 & 1.2-1.5 \\ \cline{2-7}
     & ${f_2}^0$  & 1.124  & 21.561 & 78.439 & 0.0  & 0.980 $\pm$ 0.010 \\ \cline{2-7}
     & ${f_1}^0$  & 0.274  & 78.784 & 21.211 & 0.005 & 0.4-1.2 \\ \hline
\end{tabular}
\caption{Mass spectra and components for the scalar mesons based on our fit are shown where the best value of $m_{\pi{^\prime}}$ is found to be $m_{\pi^\prime} = 1.2$ GeV.}
\label{chi1-sc}
\end{table}


In Figure \ref{figf}, we demonstrate the dependence of $\lambda_Y$ and $f_0$ masses upon the free parameter $m_Y^2$ with the rest of parameters are given in Table (\ref{chi1-prm}). As shown in Fig. (\ref{figf}a), when $m_Y^2$ is larger than $3.4$ GeV$^2$, the $\lambda_Y$ becomes negative. Then the potential of our model has to confront with the problem of unbounded vacuum from below. In the allowed values of $m_Y^2$, the masses of $f^0_i\,,i=1,2,3,4$ are almost independent of its value, as demonstrated in Fig. (\ref{figf}b). The upper bound of $m_Y^2$ is determined from the condition $\Gamma_{f^0_1} > 0.35$ GeV.



\section{Discussion and Conclusion}

In this work we develop a consistent model for the
scalar mesons below $2\,\, {\rm GeV}$ and focus on the mixing effects to the mass spectra. 
In our model we have taken into
account of the quarkonia, tetraquark and glueball scalar and
pseudo-scalar fields. Bare fields with the same quantum numbers are
allowed to mixed with each other to form the physical mesons. In
this way our isospin triplet and doublet mesons are composed of
quarkonia and tetraquark states and the isosinglet mesons are
composed with all the three chiral fields. We have presented our
prediction from the model for the scalar mass spectra on the basis
of two $\chi^2$ methods and found they yield the similar results. 

We also investigated the candidates of the glueball dominant states in our 
model. What is more encouraging is the determined value of the bare glueball mass which is
treated as a scanning parameter in our study, which is in 
agreement with the Lattice result \cite{lattice} quite well. The
consequence of the uncertainty in the bare glueball mass is also
discussed in our work.

When the constraint for $\theta_\pi$ and $\theta_K$ to vary from $\{-\frac{\pi}{4}, \frac{\pi}{4}\}$ is changed to $\{-\frac{\pi}{2}, \frac{\pi}{2}\}$, we find solutions with $m_{\Phi^\prime}^2<0$ and $m_\Phi^2>0$ which can accommodate data quite well but is in contradiction to the general belief that Pion and Kaon are quarkonia states. It is also found that when the constraint for $\lambda_Y>0$ is loosen, we can find solutions that the lightest scalar can be glueball dominant. 

In order to develop this model, we have
dropped quite a few terms and make some assumption in order to carry
out the numerical computation. We can extend current study to
those cases where the interaction terms between glueball and tetraquark field are included.
There are more than one choices available to define the interaction
between different fields, for example, our choice of the instanton
induced term is different from that in ref.~\cite{fariborz1}. It would be interesting to show the difference
between these two different parameterizations. We included the decay widths of the
$f_0(600)\rightarrow \pi \pi$ to constrain our parameter space, albeit to put a tight constraint on
our parameter sets we can consider more decay widths of all scalars and pseudo-scalars and even should include $\pi\pi $ and $\pi K $ scattering data. 
In order to examine whether our model can accommodate 
all experimental data, a global fit by treating all free parameters on the 
same footing in our model is necessary. To extend our model by including the tetraquark field as 
demonstrated in the references \cite{Forkel:2011ax,Arriola:2011en} 
to AdS/QCD framework is also straightforward. 
Following the previous study \cite{Harada:2009nq,Schaefer:2008hk,Yamamoto:2007jn,Seel:2011ju}, we can extend our model to study the role played by tetraquark states in the chiral phase transition at finite temperature and finite chemical potential.

\noindent
{\bf Acknowledgements}\\
We thank valuable discussions with F. Giacosa, T. Hatsuda,
D. Rischke and H. Q. Zheng. This work is supported by the NSFC under Grants
No. 11175251, CAS fellowship for young foreign scientists under
Grant No. 2011Y2JB05, CAS key project KJCX2-EW-N01, K.C.Wong
Education Foundation, and CAS program "Outstanding young scientists
abroad brought-in", Youth Innovation Promotion Association of CAS.

\appendix*
\section{Expression for scalar, pesudo-scalar mass matrix and decay widths}


Different elements of the scalar mass matrix:
\begin{align}
({M_s}^2)_{11} &=
 {m_\Phi}^2 + \left(v_0^2 + v_8^2\right)\lambda_1
+ \frac{1}{6} \left({v_0}^{\prime 2} + {v_8}^{\prime 2}
\right)\lambda_2
+ 2 \sqrt{\frac{2}{3}} {v_0}^{\prime } \lambda_3
- \frac{v_0 v_y}{\sqrt{3}} k   \\
({M_s}^2)_{22} &=
 {m_\Phi}^2 + \left(v_0^2 -\sqrt{2}v_0 v_8 +
\frac{3}{2} v_8^2 \right) \lambda_1
+ \left( \frac{1}{6} {v_0}^{\prime 2}  -\frac{1}{3 \sqrt{2}}
{v_0}^{\prime } {v_8}^{\prime }
+ \frac{1}{4} {v_8}^{\prime 2} \right) \lambda_2 \notag \\
&-\frac{2}{\sqrt{3}} \left(\frac{{v_0}^{\prime }}{\sqrt{2}}
+ {v_8}^{\prime } \right)\lambda_3
+ \frac{1}{\sqrt{3}} \left(\frac{v_0 v_y}{2}
+\frac{v_8 v_y}{\sqrt{2}} \right)k \\
({M_s}^2)_{33} &=
{m_{\Phi^{\prime}}}^2 + ({v_0}^{\prime 2} + {v_8}^{\prime 2}) {\lambda_1}^\prime
+\frac{1}{6}(v_0^2 + v_8^2)\lambda_2 \\
({M_s}^2)_{44} &=
 {m_{\Phi^{\prime}}}^2 + \left({v_0}^{\prime 2}
-\sqrt{2}{v_0}^{\prime } {v_8}^{\prime }
+ \frac{3 {v_8}^{\prime 2} }{2} \right){\lambda_1}^\prime
+ \left(\frac{v_0^2}{6}-\frac{v_0 v_8}{3 \sqrt{2}}
+\frac{v_8^2 }{4} \right) \lambda_2 \\
({M_s}^2)_{55} &=
{m_Y}^2 + \left(3 {v_y}^2 \right)\lambda_Y \\
({M_s}^2)_{12} &=
\frac{1}{2} \Bigl[  2(2 v_0 v_8
-\frac{v_8^2}{\sqrt{2}} )\lambda_1
+2 (\frac{{v_0}^{\prime } {v_8}^{\prime }}{3}
-\frac{{v_8}^{\prime 2}}{6 \sqrt{2}} )\lambda_2
-2(\sqrt{\frac{2}{3}}{v_8}^{\prime } )\lambda_3
\notag \\
&+(\frac{v_8 v_y}{\sqrt{3}})k  \Bigr]
 =({M_s}^2)_{21} \\
({M_s}^2)_{13} &=
\frac{1}{2} \Bigl[\lambda_m +\frac{2}{3} (v_0 {v_0}^{\prime }
+v_8 {v_8}^{\prime } )\lambda_2
+  4 \sqrt{\frac{2}{3}} v_0 \lambda_3 \Bigr] =({M_s}^2)_{31} \\
({M_s}^2)_{14} &=
\frac{1}{2} \Bigl[ \frac{2}{3} (v_8 {v_0}^{\prime }
+v_0 {v_8}^{\prime }
-\frac{v_8 {v_8}^{\prime }}{\sqrt{2}})\lambda_2
-2(\sqrt{\frac{2}{3}} v_8 )\lambda_3 \Bigr]  =({M_s}^2)_{41} \\
({M_s}^2)_{15} &=
\frac{1}{2} \Bigl[ -\frac{1}{\sqrt{3}} (v_0^2
- \frac{v_8^2}{2})k  \Bigr] =({M_s}^2)_{51} \\
({M_s}^2)_{23} &=
\frac{1}{2} \Bigl[ \frac{2}{3}(v_8 {v_0}^{\prime }
+ v_0 {v_8}^{\prime }
-\frac{v_8 {v_8}^{\prime }}{\sqrt{2}} )\lambda_2
-2 \sqrt{\frac{2}{3}} v_8 \lambda_3  \Bigr] =({M_s}^2)_{32} \\
({M_s}^2)_{24} &=
\frac{1}{2} \Bigl[\lambda_m + 2( \frac{v_0 {v_0}^{\prime }}{3}
-\frac{v_8 {v_0}^{\prime }}{3\sqrt{2}}
-\frac{v_0 {v_8}^{\prime }}{3\sqrt{2}}
+\frac{v_8 {v_8}^{\prime }}{2}  )\lambda_2 \notag \\
&-\frac{2\sqrt{2}}{\sqrt{3}} (v_0 + \sqrt{2} v_8
)\lambda_3   \Bigr] =({M_s}^2)_{42} \\
({M_s}^2)_{25} &=
\frac{1}{2} \Bigl[ \frac{1}{\sqrt{3}} (v_0 v_8
+ \frac{ {v_8}^2}{\sqrt{2}})k \Bigr] =({M_s}^2)_{52} \\
({M_s}^2)_{34} &=
\frac{1}{2} \Bigl[ 2(2 {v_0}^{\prime }{v_8}^{\prime }
-\frac{{v_8}^{\prime 2}}{\sqrt{2}} ){\lambda_1}^\prime
+\frac{2}{3}(v_0 v_8 -\frac{{v_8}^2}{2 \sqrt{2}}
)\lambda_2 \Bigr] =({M_s}^2)_{43} \\
({M_s}^2)_{35} &= ({M_s}^2)_{53} =({M_s}^2)_{45} = ({M_s}^2)_{54} = 0
\end{align}

where, ${M^2}_{\sigma_0 \sigma_0} = ({M_s}^2)_{11}$,
    ${M^2}_{\sigma_8 \sigma_8} = ({M_s}^2)_{22}$,
    ${M^2}_{{\sigma_0}^\prime {\sigma_0}^\prime} = ({M_s}^2)_{33}$,
    ${M^2}_{{\sigma_8}^\prime {\sigma_8}^\prime} = ({M_s}^2)_{44}$,
    ${M^2}_{y_1 y_1} = ({M_s}^2)_{55}$,
    ${M^2}_{\sigma_0 \sigma_8} = ({M_s}^2)_{12}$,
    ${M^2}_{\sigma_0 {\sigma_0}^\prime} = ({M_s}^2)_{13}$,
    ${M^2}_{\sigma_0 {\sigma_8}^\prime} = ({M_s}^2)_{14}$,
    ${M^2}_{\sigma_0 y_1} = ({M_s}^2)_{15}$,
    ${M^2}_{\sigma_8 {\sigma_0}^\prime} = ({M_s}^2)_{23}$,
    ${M^2}_{\sigma_8 {\sigma_8}^\prime} = ({M_s}^2)_{24}$,
    ${M^2}_{\sigma_8 y_1} = ({M_s}^2)_{25}$,
    ${M^2}_{{\sigma_0}^\prime {\sigma_8}^\prime} = ({M_s}^2)_{34}$,
    ${M^2}_{{\sigma_0}^\prime y_1} = ({M_s}^2)_{35}$,
    ${M^2}_{{\sigma_8}^\prime y_1} = ({M_s}^2)_{45}$.

Different elements of the pseudo-scalar mass matrix:

\begin{align}
({M_\eta}^2)_{11} &=
 {m_\Phi}^2 + \frac{1}{3} \Bigl(v_0^2
+ v_8^2\Bigr)\lambda_1
+ \frac{1}{6} \Bigl( {v_0}^{\prime 2} + {v_8}^{\prime 2}
\Bigr)\lambda_2
-2\Bigl(\sqrt{\frac{2}{3}}{v_0}^{\prime } \Bigr)\lambda_3
+\Bigl( \frac{v_0 v_y}{\sqrt{3}}\Bigr)k   \\
({M_\eta}^2)_{22} &=
 {m_\Phi}^2 + 2\Bigl(\frac{v_0^2}{6} -\frac{v_0 v_8}
{3 \sqrt{2}} +\frac{v_8^2}{4}\Bigr)\lambda_1
+ \Bigl(\frac{{v_0}^{\prime 2}}{6}
-\frac{{v_0}^{\prime }{v_8}^{\prime }}{3 \sqrt{2}}
+\frac{{v_8}^{\prime 2}}{4}\Bigr)\lambda_2 \notag \\
&+ \frac{2}{\sqrt{3}}\Bigl( \frac{{v_0}^{\prime }}{\sqrt{2}}
+{v_8}^{\prime }\Bigr)\lambda_3
- \frac{1}{\sqrt{3}}\Bigl( \frac{v_0 v_y}{2}
+ \frac{v_8 v_y}{\sqrt{2}}\Bigr)k   \\
({M_\eta}^2)_{33} &=
 {m_{\Phi^{\prime}}}^2 +\frac{1}{3}\Bigl({v_0}^{\prime 2}
+ {v_8}^{\prime 2}\Bigr){\lambda_1}^\prime
+\frac{1}{6}\Bigl(v_0^2 + v_8^2\Bigr)\lambda_2  \\
({M_\eta}^2)_{44} &=
 {m_{\Phi^{\prime}}}^2 +2 \Bigl(\frac{{v_0}^{\prime 2}}{6}
-\frac{{v_0}^{\prime }{v_8}^{\prime }}{3 \sqrt{2}}
+\frac{{v_8}^{\prime 2}}{4}\Bigr){\lambda_1}^\prime
+\Bigl(\frac{v_0^2}{6} -\frac{v_0 v_8 }{3 \sqrt{2}}
+\frac{v_8^2}{4}\Bigr)\lambda_2  \\
({M_\eta}^2)_{55} &=
{m_Y}^2 + {v_y}^2 \lambda_Y \\
({M_\eta}^2)_{12} &=
\frac{1}{2} \Bigl[ \frac{2}{3}\Bigl(2 v_0 v_8
-\frac{v_8^2}{\sqrt{2}} \Bigr)\lambda_1
+ \frac{2}{3}\Bigl({v_0}^{\prime }{v_8}^{\prime }
-\frac{{v_8}^{\prime 2}}{2 \sqrt{2}} \Bigr)\lambda_2
+2\Bigl(\sqrt{\frac{2}{3}} {v_8}^{\prime } \Bigr)\lambda_3 \notag \\
&-\Bigl(\frac{v_8 v_y}{\sqrt{3}}\Bigr)k  \Bigr] = ({M_\eta}^2)_{21} \\
({M_\eta}^2)_{13} &=
\frac{1}{2} \Bigl[\lambda_m -4 \sqrt{\frac{2}{3}} v_0 \lambda_3 \Bigr] =({M_\eta}^2)_{31} \\
({M_\eta}^2)_{14} &=
\frac{1}{2} \Bigl[ 2 \sqrt{\frac{2}{3}} v_8 \lambda_3 \Bigr] =({M_\eta}^2)_{41}\\
({M_\eta}^2)_{15} &=
\frac{1}{2} \Bigl[ \frac{1}{\sqrt{3}} \Bigl(v_0^2
- \frac{{v_8}^2}{2 \sqrt{3}}\Bigr)k  \Bigr] = ({M_\eta}^2)_{51} \\
({M_\eta}^2)_{23} &=
\frac{1}{2} \Bigl[2 \sqrt{\frac{2}{3}} v_8 \lambda_3 \Bigr] = ({M_\eta}^2)_{32} \\
({M_\eta}^2)_{24} &=
\frac{1}{2} \Bigl[\lambda_m + 2 \Bigl(\sqrt{\frac{2}{3}} v_0 + \frac{2}{\sqrt{3}}
v_8 \Bigr) \lambda_3  \Bigr] = ({M_\eta}^2)_{42} \\
({M_\eta}^2)_{25} &=
\frac{1}{2} \Bigl[-\frac{1}{\sqrt{3}} \Bigl(v_0 v_8
+ \frac{v_8^2}{\sqrt{2}} \Bigr)k \Bigr] = ({M_\eta}^2)_{52} \\
({M_\eta}^2)_{34} &=
\frac{1}{2} \Bigl[  \frac{2}{3} \Bigl( 2 {v_0}^{\prime }{v_8}^{\prime }
- \frac{{v_8}^{\prime 2}}{\sqrt{2}} \Bigr){\lambda_1}^\prime
+\frac{2}{3} \Bigl(v_0 v_8 - \frac{v_8^2}{3 \sqrt{2}}
\Bigr)\lambda_2 \Bigr] = ({M_\eta}^2)_{43} \\
({M_\eta}^2)_{35} &= ({M_\eta}^2)_{53} =({M_\eta}^2)_{45} = ({M_\eta}^2)_{54} = 0
\end{align}

where, ${M^2}_{\pi_0 \pi_0} = ({M_\eta}^2)_{11} $
    ${M^2}_{\pi_8 \pi_8} = ({M_\eta}^2)_{22} $
    ${M^2}_{{\pi_0}^\prime {\pi_0}^\prime} = ({M_\eta}^2)_{33} $
    ${M^2}_{{\pi_8}^\prime {\pi_8}^\prime} = ({M_\eta}^2)_{44} $
    ${M^2}_{y_2 y_2} = ({M_\eta}^2)_{55} $
    ${M^2}_{\pi_0 \pi_8} = ({M_\eta}^2)_{12} $
    ${M^2}_{\pi_0 {\pi_0}^\prime} = ({M_\eta}^2)_{13} $
    ${M^2}_{\pi_0 {\pi_8}^\prime} = ({M_\eta}^2)_{14} $
    ${M^2}_{\pi_0 y_1} = ({M_\eta}^2)_{15} $
    ${M^2}_{\pi_8 {\pi_0}^\prime} = ({M_\eta}^2)_{23} $
    ${M^2}_{\pi_8 {\pi_8}^\prime} = ({M_\eta}^2)_{24} $
    ${M^2}_{\pi_8 y_1} = ({M_\eta}^2)_{25} $
    ${M^2}_{{\pi_0}^\prime {\pi_8}^\prime} = ({M_\eta}^2)_{34} $
    ${M^2}_{{\pi_0}^\prime y_2} = ({M_\eta}^2)_{35} $
    ${M^2}_{{\pi_8}^\prime y_2} = ({M_\eta}^2)_{45} $.

\vskip 0.5cm

For the decay constant, we have taken the following standard formula:
corresponding to the interaction Lagrangian
${\cal{L}}_{int} =G f_0 \pi_p \pi_p$ (the subscript 'p' denotes the physical
pion fields), the decay constant is given by:

\begin{align}
\Gamma = 3 s_f \frac{k_f}{8 \pi {m_{f_0}}^2} \big|-i M\big|^2
\end{align}

Where, $s_f$ is the symmetry factor, which is in our case is $\frac{1}{2}$
and $k_f=\sqrt{ \frac{{m_{f_0}}^2}{4} - {m_{\pi_p}}^2}$. At the tree level
$\big|-i M \big|^2 = G^2$. We calculated this coupling constant from our
bare Lagrangian following the procedure presented in \cite{fariborz3}
The explicit expression for the coupling constant is given below
(where $R_s$ stands for the rotation mass matrix for scalars):

\begin{center}
\begin{align}
g_{11} &=-\sqrt{2} \bigl( \sqrt{2} v_0 + v_8 \bigr) \lambda_1
- \frac{v_y k}{2 \sqrt{3}} \\ \notag
g_{12} &=- \frac{2}{3} \bigl( {v_0}^\prime + \frac{{v_8}^\prime}{\sqrt{2}} \bigr)
\lambda_2 + 2 \sqrt{\frac{2}{3}} \lambda_3 \\ \notag
g_{13} &= -\frac{1}{3} \bigl( v_0 + \frac{v_8}{\sqrt{2}} \bigr) \lambda_2 \\ \notag
g_{21} &= -\bigl( \sqrt{2} v_0 + v_8 \bigr) \lambda_1
+ \frac{v_y k}{\sqrt{6}} \\ \notag
g_{22} &=- \frac{1}{3} \bigl( \sqrt{2} {v_0}^\prime + {v_8}^\prime \bigr)
\lambda_2 - \frac{4 \lambda_3}{\sqrt{3}} \\ \notag
g_{23} &= -\frac{1}{3\sqrt{2}} \bigl( v_0 + \frac{v_8}{\sqrt{2}} \bigr)
\lambda_2 \\ \notag
g_{31} &=- \frac{1}{3} \bigl( {v_0}^\prime + \frac{{v_8}^\prime}{\sqrt{2}}
\bigr) \lambda_2 + \sqrt{\frac{2}{3}} \lambda_3 \\ \notag
g_{32} &=-\frac{2}{3} \bigl( v_0 + \frac{v_8}{\sqrt{2}} \bigr) \lambda_2 \\ \notag
g_{33} &= - \sqrt{2} \bigl( \sqrt{2} {v_0}^\prime + {v_8}^\prime \bigr)
{\lambda_1}^\prime \\ \notag
g_{41} &=- \frac{1}{3 \sqrt{2}} \bigl( {v_0}^\prime +
\frac{{v_8}^\prime}{\sqrt{2}} \bigr) \lambda_2 -
\frac{2}{\sqrt{3}} \lambda_3 \\ \notag
g_{42} &=- \frac{1}{3} \bigl( \sqrt{2} v_0 +  v_8 \bigr) \lambda_2 \\ \notag
g_{43} &= - \bigl( \sqrt{2} {v_0}^\prime + {v_8}^\prime \bigr)
{\lambda_1}^\prime \\ \notag
g_{51} &=-\frac{1}{\sqrt{6}} \bigl( \frac{v_0}{\sqrt{2}}
-v_8 \bigr) k \\ \notag
g_{52} & = 0 = g_{53} \\ \notag
G_1  &= \bigl( R_s \bigr)_{51} \bigl[ {\bigl(R_{\pi \pi^\prime} \bigr)_{11}}^2
g_{11} + \bigl(R_{\pi \pi^\prime} \bigr)_{11} \bigl(R_{\pi \pi^\prime} \bigr)_{12} g_{12}
+{\bigl(R_{\pi \pi^\prime} \bigr)_{12}}^2 g_{13}  \bigr] \\ \notag
G_2  &= \bigl( R_s \bigr)_{52} \bigl[ {\bigl(R_{\pi \pi^\prime} \bigr)_{11}}^2
g_{21} + \bigl(R_{\pi \pi^\prime} \bigr)_{11} \bigl(R_{\pi \pi^\prime} \bigr)_{12} g_{22}
+{\bigl(R_{\pi \pi^\prime} \bigr)_{12}}^2 g_{23}  \bigr] \\ \notag
G_3  &= \bigl( R_s \bigr)_{53} \bigl[ {\bigl(R_{\pi \pi^\prime} \bigr)_{11}}^2
g_{31} + \bigl(R_{\pi \pi^\prime} \bigr)_{11} \bigl(R_{\pi \pi^\prime} \bigr)_{12} g_{32}
+{\bigl(R_{\pi \pi^\prime} \bigr)_{12}}^2 g_{33}  \bigr] \\ \notag
G_4  &= \bigl( R_s \bigr)_{54} \bigl[ {\bigl(R_{\pi \pi^\prime} \bigr)_{11}}^2
g_{41} + \bigl(R_{\pi \pi^\prime} \bigr)_{11} \bigl(R_{\pi \pi^\prime} \bigr)_{12} g_{42}
+{\bigl(R_{\pi \pi^\prime} \bigr)_{12}}^2 g_{43}  \bigr] \\ \notag
G_5  &= \bigl( R_s \bigr)_{55} \bigl[ {\bigl(R_{\pi \pi^\prime} \bigr)_{11}}^2
g_{51} + \bigl(R_{\pi \pi^\prime} \bigr)_{11} \bigl(R_{\pi \pi^\prime} \bigr)_{12} g_{52}
+{\bigl(R_{\pi \pi^\prime} \bigr)_{12}}^2 g_{53}  \bigr] \\ \notag
G  &= G_1 + G_2 + G_3 + G_4 + G_5
\end{align}
\end{center}

The expressions of mass matrices for $a-a^\prime$ and $\pi-\pi^\prime$
mesons are given below:

\begin{center}
\begin{align}
\bigl( {M_{aa^\prime}}^2 \bigr)_{11} &= {m_\Phi}^2 - A_{11} \lambda_1
- B_{11} \lambda_2 -C_{11} \lambda_3 -D_{11} v_y k \\
\bigl( {M_{aa^\prime}}^2 \bigr)_{22} &= {m_{\Phi^\prime}}^2
- A_{22} {\lambda_1}^\prime -B_{22} \lambda_2 \\
\bigl( {M_{aa^\prime}}^2 \bigr)_{12} &= \frac{1}{2} \bigl[ \lambda_m
- A_{12} \lambda_2 - B_{12} \lambda_3 \bigr] \\
\bigl( {M_{\pi \pi^\prime}}^2 \bigr)_{11} &= {m_\Phi}^2 - \frac{1}{3}
A_{11} \lambda_1 - B_{11} \lambda_2 +C_{11} \lambda_3 +D_{11} v_y k \\
\bigl( {M_{\pi \pi^\prime}}^2 \bigr)_{22} &= {m_{\Phi^\prime}}^2
-\frac{1}{3} A_{22} {\lambda_1}^\prime -B_{22} \lambda_2 \\
\bigl( {M_{\pi \pi^\prime}}^2 \bigr)_{12} &= \frac{1}{2} \bigl[ \lambda_m
+ B_{12} \lambda_3 \bigr] \\
\text{Where,} \nonumber \\
A_{11} &= -{v_0}^2 - \sqrt{2} v_0 v_8 - \frac{{v_8}^2}{2} \\
B_{11} &= -\frac{{v_0}^{\prime 2}}{6} - \frac{{v_0}^\prime {v_8}^\prime}{3 \sqrt{2}}
-\frac{{v_8}^{\prime 2}}{12} \\
C_{11} &= \sqrt{\frac{2}{3}} {v_0}^\prime -\frac{2}{\sqrt{3}} {v_8}^\prime \\
D_{11} &= -\frac{v_0}{2 \sqrt{3}} + \frac{v_8}{\sqrt{6}} \\
A_{22} &=-{v_0}^{\prime 2} -\sqrt{2} {v_0}^ \prime {v_8}^\prime
-\frac{{v_8}^{\prime 2}}{2} \\
B_{22} &= -\frac{{v_0}^2}{6} - \frac{v_0 v_8}{3 \sqrt{2}}
-\frac{{v_8}^2}{12} \\
A_{12} &= - \frac{2}{3} v_0 {v_0}^\prime -\frac{\sqrt{2}}{3}
v_8 {v_0}^\prime - \frac{\sqrt{2}}{3} v_0 {v_8}^\prime
- \frac{1}{3} v_8 {v_8}^\prime \\
B_{12} &= 2 \sqrt{\frac{2}{3}} v_0 - \frac{4}{\sqrt{3}} v_8
\end{align}
\end{center}

The expressions of mass matrices for $\kappa-\kappa^\prime$ and $K-K^\prime$
mesons are given below:

\begin{center}
\begin{align}
\bigl( {M_{\kappa \kappa^\prime}}^2 \bigr)_{11} &= {m_\Phi}^2 - E_{11} \lambda_1
- F_{11} \lambda_2 -G_{11} \lambda_3 -H_{11} v_y k \\
\bigl( {M_{\kappa \kappa^\prime}}^2 \bigr)_{22} &= {m_{\Phi^\prime}}^2
- E_{22} {\lambda_1}^\prime -F_{22} \lambda_2 \\
\bigl( {M_{\kappa \kappa^\prime}}^2 \bigr)_{12} &= \frac{1}{2} \bigl[ \lambda_m
- E_{12} \lambda_2 - F_{12} \lambda_3 \bigr] \\
\bigl( {M_{KK^\prime}}^2 \bigr)_{11} &= {m_\Phi}^2 -
I_{11} \lambda_1 - F_{11} \lambda_2 +G_{11} \lambda_3 +H_{11} v_y k \\
\bigl( {M_{KK^\prime}}^2 \bigr)_{22} &= {m_{\Phi^\prime}}^2
- J_{22} {\lambda_1}^\prime -F_{22} \lambda_2 \\
\bigl( {M_{KK^\prime}}^2 \bigr)_{12} &= \frac{1}{2} \bigl[ \lambda_m
- K_{12} \lambda_2 + F_{12} \lambda_3 \bigr] \\
\text{Where,}  \notag \\
E_{11} &= -{v_0}^2 + \frac{v_0 v_8}{\sqrt{2}}  - \frac{{v_8}^2}{2} \\
F_{11} &= -\frac{{v_0}^{\prime 2}}{6} + \frac{{v_0}^\prime {v_8}^\prime}{6 \sqrt{2}}
-\frac{5 {v_8}^{\prime 2}}{24} \\
G_{11} &= \sqrt{\frac{2}{3}} {v_0}^\prime + \frac{{v_8}^\prime}{\sqrt{3}}  \\
H_{11} &= -\frac{v_0}{2 \sqrt{3}} - \frac{v_8}{2\sqrt{6}} \\
E_{22} &=-{v_0}^{\prime 2} +\frac{{v_0}^ \prime {v_8}^\prime}{\sqrt{2}}
-\frac{{v_8}^{\prime 2}}{2} \\
F_{22} &= -\frac{{v_0}^2}{6} + \frac{v_0 v_8}{6 \sqrt{2}}
-\frac{5 {v_8}^2}{24} \\
E_{12} &= - \frac{2}{3} v_0 {v_0}^\prime +\frac{v_8 {v_0}^\prime}{3 \sqrt{2}}
 + \frac{v_0 {v_8}^\prime}{3\sqrt{2}}
- \frac{1}{12} v_8 {v_8}^\prime \\
F_{12} &= 2 \sqrt{\frac{2}{3}} v_0 + \frac{2}{\sqrt{3}} v_8 \\
I_{11} &= -\frac{{v_0}^2}{3} + \frac{v_0 v_8}{3 \sqrt{2}}
-\frac{7 {v_8}^2}{6} \\
J_{22} &= -\frac{{v_0}^{\prime 2}}{3} + \frac{{v_0}^\prime {v_8}^\prime}{3 \sqrt{2}}
-\frac{7 {v_8}^{\prime 2}}{6} \\
K_{12} &= - \frac{3}{4} v_8 {v_8}^\prime
\end{align}
\end{center}


\begin{thebibliography}{9}

\bibitem{pdg2010}
K. Nakamura et al. (Particle Data Group), J. \ Phys. \ {\bf G37}, 075021 (2010) 

\bibitem{Scalar}
K.~F.~Liu,
  arXiv:0805.3364 [hep-lat];
K.~F.~Liu and C.~W.~Wong,
  Phys.\ Lett.\  B {\bf 107}, 391 (1981);
H.~Y.~Cheng, C.~K.~Chua and K.~F.~Liu,
  Phys.\ Rev.\  D {\bf 74}, 094005 (2006);
G.~'.~Hooft, G.~Isidori, L.~Maiani, A.~D.~Polosa and V.~Riquer,
  Phys.\ Lett.\  B {\bf 662}, 424 (2008);
Q.~Zhao, B.~s.~Zou and Z.~b.~Ma,
  Phys.\ Lett.\  B {\bf 631}, 22 (2005);
D.~V.~Bugg, M.~J.~Peardon and B.~S.~Zou,
  Phys.\ Lett.\  B {\bf 486}, 49 (2000).

\bibitem{Scalar-lightGB}S.~Narison,
  Nucl.\ Phys.\ Proc.\ Suppl.\  {\bf 186}, 306 (2009)

\bibitem{Glueball-Review}
 V.~Mathieu, N.~Kochelev and V.~Vento,
  Int.\ J.\ Mod.\ Phys.\  E {\bf 18}, 1 (2009);
E.~Klempt and A.~Zaitsev,
  Phys.\ Rept.\  {\bf 454}, 1 (2007);
C.~Amsler and N.~A.~Tornqvist,
  Phys.\ Rept.\  {\bf 389}, 61 (2004).

\bibitem{Caprini:2005zr} 
  I.~Caprini, G.~Colangelo and H.~Leutwyler,
  Phys.\ Rev.\ Lett.\  {\bf 96}, 132001 (2006)
  [hep-ph/0512364].

\bibitem{Black:1998wt} 
  D.~Black, A.~H.~Fariborz, F.~Sannino and J.~Schechter,
  Phys.\ Rev.\ D {\bf 59}, 074026 (1999)
  [hep-ph/9808415].
  

\bibitem{Bugg:2009uk} 
  D.~V.~Bugg,
  Phys.\ Rev.\ D {\bf 81}, 014002 (2010)
  [arXiv:0906.3992 [hep-ph]].
  
\bibitem{Ablikim:2005ni} 
  M.~Ablikim {\it et al.}  [BES Collaboration],
  Phys.\ Lett.\ B {\bf 633}, 681 (2006)
  [hep-ex/0506055].

\bibitem{jaffe} R. L Jaffe, Phys. Rev. D15, 267, 1977.

\bibitem{Achasov:2009ee}
  N.~N.~Achasov and G.~N.~Shestakov,
  Phys.\ Usp.\  {\bf 54}, 799 (2011)
  [arXiv:0905.2017 [hep-ph]].

\bibitem{molecule}J. D. Weinstein and N. Isgur, Phys. Rev. Lett 48, 659, 1982.

\bibitem{Minkowski:1998mf} 
  P.~Minkowski and W.~Ochs,
  Eur.\ Phys.\ J.\ C {\bf 9}, 283 (1999)
  [hep-ph/9811518].

\bibitem{arXiv:0804.4452}
  G.~Mennessier, S.~Narison and W.~Ochs,
  Phys.\ Lett.\ B\ {\bf 665}, 205  (2008)
  [arXiv:0804.4452 [hep-ph]].

\bibitem{Mennessier:2010xg}
  G.~Mennessier, S.~Narison and X.~G.~Wang,
  Phys.\ Lett.\ B {\bf 688}, 59 (2010)
  [arXiv:1002.1402 [hep-ph]].

\bibitem{Dai:2011bs}
  L.~Y.~Dai, X.~G.~Wang and H.~Q.~Zheng,
  arXiv:1108.1451 [hep-ph].

\bibitem{Alford:2000mm} 
  M.~G.~Alford and R.~L.~Jaffe,
  Nucl.\ Phys.\ B {\bf 578}, 367 (2000)
  [hep-lat/0001023].
  
\bibitem{Prelovsek:2010kg} 
  S.~Prelovsek, T.~Draper, C.~B.~Lang, M.~Limmer, K.~-F.~Liu, N.~Mathur and D.~Mohler,
  Phys.\ Rev.\ D {\bf 82}, 094507 (2010)
  [arXiv:1005.0948 [hep-lat]].

\bibitem{qishu} S. He, M. Huang and Q. S. Yan, Phys. Rev. D 81, 014003 (2010).

\bibitem{Fariborz:2011es}
  A.~H.~Fariborz,
  arXiv:1109.2630 [hep-ph].

\bibitem{Mathieu:2008me}
  V.~Mathieu, N.~Kochelev and V.~Vento,
  Int.\ J.\ Mod.\ Phys.\ E {\bf 18}, 1 (2009)
  [arXiv:0810.4453 [hep-ph]].

\bibitem{u11}G. Hooft, Phys. Rep. 142, 357 (1986).

\bibitem{fariborz1}A. H. Fariborz, R. Jora, J. Schechter, Phys. Rev. D 76, 014011 (2007).

\bibitem{Schaefer:2008hk} 
  B.~-J.~Schaefer and M.~Wagner,
  Phys.\ Rev.\ D {\bf 79}, 014018 (2009)
  [arXiv:0808.1491 [hep-ph]].
  
\bibitem{Lenaghan:2000ey} 
  J.~T.~Lenaghan, D.~H.~Rischke and J.~Schaffner-Bielich,
  Phys.\ Rev.\ D {\bf 62}, 085008 (2000)
  [nucl-th/0004006].


\bibitem{fariborz2}A. H. Fariborz, R. Jora, J. Schechter, Phys. Rev. D 79, 074014 (2009).


\bibitem{giacosa} F. Giacosa, Phys. Rev. D 75, 054007 (2007).

\bibitem{lattice} C. Michael, Hadron 97 Conference, AIP Conf. Proc. 432 (1998) 657.

\bibitem{Forkel:2011ax} 
  H.~Forkel,
  AIP Conf.\ Proc.\  {\bf 1388}, 182 (2011)
  [arXiv:1103.3902 [hep-ph]].
  
\bibitem{Arriola:2011en} 
  E.~R.~Arriola and W.~Broniowski,
  (Bled Workshops in Physics. Vol. 12 No. 1)
  [arXiv:1110.2863 [hep-ph]].

\bibitem{fariborz3}A. H. Fariborz, R. Jora, J. Schechter and M. N. Sahid
arXiv:1106.4538 [hep-ph].

\bibitem{Harada:2009nq} 
  M.~Harada, C.~Sasaki and S.~Takemoto,
  Phys.\ Rev.\ D {\bf 81}, 016009 (2010)
  [arXiv:0908.1361 [hep-ph]].

\bibitem{Yamamoto:2007jn} 
  N.~Yamamoto, T.~Hatsuda, M.~Tachibana and G.~Baym,
  J.\ Phys.\ G G {\bf 34}, S635 (2007)
  [hep-ph/0701191].

\bibitem{Seel:2011ju} 
  E.~Seel, S.~Struber, F.~Giacosa and D.~H.~Rischke,
  arXiv:1108.1918 [hep-ph].
    
\end{thebibliography}
\end{document}